\documentclass[twocolumn,prb,aps,showpacs,amsmath]{revtex4}
\newcommand{\BEQ}{\begin{equation}}     % Gleichungen Anfang ..
\newcommand{\BEA}{\begin{eqnarray}}
\newcommand{\EEQ}{\end{equation}}       % .. und Ende
\newcommand{\EEA}{\end{eqnarray}}
\newcommand{\eps}{\epsilon}          % epsilon
\newcommand{\D}{{\rm d}}
\begin{document}

\input epsf.sty

\title{Corrections to local scale invariance in the non-equilibrium
dynamics of critical systems: numerical evidences}

\author{Michel Pleimling$^1$ and Andrea Gambassi$^{2,3}$}

\affiliation{$^1$~Institut f\"ur Theoretische Physik I, Universit\"at Erlangen-N\"urnberg,
D -- 91058 Erlangen, Germany\\
$^2$~Max-Planck Institut f\"ur Metallforschung, 
Heisenbergstr. 3, D -- 70569 Stuttgart, Germany\\
$^3$~Institut f\"ur Theoretische und Angewandte Physik,
Universit\"at Stuttgart, Pfaffenwaldring 57, D -- 70569 Stuttgart, Germany
}

\begin{abstract}
Local scale invariance (LSI) 
has been recently proposed as a possible extension of
the dynamical scaling in systems at the critical point and during
phase ordering. 
LSI has been applied {\it inter alia} to provide predictions for
the scaling properties of the response function of non-equilibrium
critical systems in the aging regime following a quench from the
high-temperature phase to the critical point.
These predictions 
have been confirmed by Monte Carlo simulations and
analytical results for some specific models, but
they are in disagreement with field-theoretical predictions. 
By means of Monte Carlo simulations of the critical
two- and three-dimensional Ising model with Glauber dynamics,
we study the intermediate integrated response, finding
deviations from the corresponding LSI predictions that are in qualitative
agreement with the field-theoretical computations.
This result casts some doubts on the general applicability of LSI to
critical dynamics.
\end{abstract}


\maketitle

\section{Introduction}
The non-equilibrium dynamics of systems at the critical point 
has been recently the subject of a renewed interest due to its
similarities with the aging phenomena occurring in a variety of glassy
materials, see Refs.~\onlinecite{Cug02,God02,Cal04b} for reviews.
In the typical scenario the system is prepared in the high-temperature
equilibrium state. Then, at time $t=0$, it is quenched to
the critical temperature $T_c$.
Due to the critical slowing down, after this sudden thermal
perturbation the system undergoes
a neverending relaxation towards the equilibrium state. 
Interesting scaling properties of this evolution are revealed by
two-time quantities such as the correlation function 
$C_{\bf x}(t,s)=\langle \phi_{\bf x}(t)\phi_{\bf 0}(s)\rangle$ 
and the linear response function
$R_{\bf x}(t,s)=\delta 
\langle \phi_{\bf x}(t)\rangle/\delta h_{\bf 0}(s)|_{h=0}$, where
$\phi_{\bf x}(t)$ is the order parameter at the point ${\bf x}$
and time $t$ and $h_{\bf 0}(s)$ is the conjugated field at time $s<t$
and position ${\bf x}= {\bf 0}$ (here translational invariance is assumed). 

In thermal equilibrium $C$ and $R$ depend
on $t-s$ and they are related by the fluctuation-dissipation theorem,
whereas in the present case they are non-trivial functions of both
times. This is the so-called aging regime.
\cite{Cug02,God02,Cal04b} 
In particular, in the dynamical scaling regime
$s\gg \tau_{\rm micro} ~~\mbox{and} ~~ t-s\gg \tau_{\rm micro}$, 
where $\tau_{\rm micro}$
is some microscopic time, the autoresponse function $R_{{\bf x}={\bf
0}}$ is expected to scale as:\cite{God02,Cug02}
\BEQ 
R_{{\bf x}={\bf 0}}(t,s) = {\mathcal A}_R (t-s)^{-(a+1)}(t/s)^\theta {\mathcal F}_R(s/t).
\label{eq:scal:autoR}
\EEQ
The scaling function ${\mathcal F}_R(v)$ 
is universal in the sense that it depends only on the universality class of the
statistical system, provided one fixes the non-universal constant
${\mathcal A}_R$ to have ${\mathcal F}_R(0)=1$. The exponent $\theta$
is related to the initial-slip exponent of the magnetization\cite{Jan89}
(equivalently to the autocorrelation exponent\cite{Hus89} $\lambda$, as $\theta =
1 + a -\lambda/z$), whereas 
$a=2\beta/(\nu z)= (d-2+\eta)/z$. 
Here $\beta$, $\nu$ and $\eta$ are the usual static critical exponents, $d$ is
the number of space dimensions and $z$ is the dynamic critical exponent.

Eq.~(\ref{eq:scal:autoR}) is
basically obtained (for example within the renormalization-group
approach\cite{Cal04b,Jan89})
by exploiting the fact that, at the critical point, 
the response function transforms covariantly under dilatations in
space ${\bf x}\mapsto b {\bf x}$ and time $t \mapsto b^z t$. 
It is possible to {\it prove}, under quite general assumptions,
that the covariance of static correlation functions (of suitable
operators) under a scale 
transformation ${\bf x}\mapsto b{\bf x}$ 
extends to conformal transformations, resulting in strong constraints 
on the functional form of the associated scaling functions.
%
%For static correlation functions (of suitable operators) the
%covariance under a scale transformation ${\bf x}\mapsto b {\bf x}$ 
%extends to conformal transformations, resulting in strong
%constraints on the functional form of the associated 
%scaling functions, in any space dimension $d$. 
%
It is then natural to address the question whether an analogous
extension carries over to the case of dynamic scaling with $z\neq 1$, where
the dilatation factors in space and time are different. (In strongly
anisotropic systems the role of the time is formally played 
by a special direction in space\cite{Ple01}.)
This possibility has been recently investigated, 
leading to the identification of a group of local scale
transformations which generalize the dynamic scaling.\cite{Hen01,Hen02}
The covariance of the dynamic correlation functions (of
suitable operators) under such a group is called Local Scale
Invariance (LSI). 
The conditions under which the dynamical scaling implies LSI have not
yet been established beyond the specific case\cite{HU03} $z=2$. 
In the absence of such a general result 
one can heuristically {\it assume} LSI and then check whether its 
predictions are actually confirmed. 
%
%The assumption that 
%LSI applies to the response function
%in aging systems (actually considering only a subgroup of the original
%group of local scale transformations) 
%yields\cite{Hen01,Hen02,footnote1}
%\BEQ \label{eq:lsi:x}
%{\mathcal F}^{\rm (LSI)}_R(v)\equiv 1\;
%\EEQ
%for the scaling function in Eq.~(\ref{eq:scal:autoR}).
For the response function in aging systems [see Eq.~(\ref{eq:scal:autoR})] LSI 
yields\cite{Hen01,Hen02,footnote1}
\BEQ \label{eq:lsi:x}
{\mathcal F}^{\rm (LSI)}_R(v)\equiv 1\;.
\EEQ

This prediction has been found 
in excellent agreement with Monte Carlo simulations of
two- and three-dimensional Ising models and three-dimensional $XY$
model, with Glauber dynamics, both at and below
$T_c$.\cite{Hen01,Hen03,Hen03a,Abr04b} In all these simulations
the measured quantity has been either the autoresponse function
$R_{{\bf x}={\bf 0}}(t,s)$ or some related integrated responses. 
Moreover the prediction~(\ref{eq:lsi:x}) 
has been also confirmed by analytical results
for spherical models with non-conserved order 
parameter\cite{Jan89,God00} and for the two-dimensional $XY$ model in the
spin-wave approximation.\cite{Pic04}
These %results 
findings
strongly support the idea that LSI is 
%indeed 
actually realized at the
critical point, 
providing 
%a useful tool to 
constrains on the form of scaling functions.
%, in analogy to what happens with conformal invariance.
%
%
%

For the spatial dependence of the response function LSI predicts\cite{Hen02}
\BEQ \label{eq:scal:Rx}
R_{\bf x}(t,s) = R_{{\bf x} = {\bf 0}}(t,s) \Phi(\left| {\bf x}\right| (t-s)^{-1/z}) \;,
\EEQ
where the function $\Phi$ satisfies a (fractional) differential
equation whose general solution is reported in Ref.~\onlinecite{Hen02}.

Instead of studying numerically local quantities,
%addition to 
%the local quantities usually studied numerically,
one can 
%also 
consider
global ones such as the time-dependent
total order parameter $M(t) = \int\D^dx
\,\phi_{\bf x}(t) = \phi_{{\bf q}={\bf 0}}(t)$ (${\bf q}$ is the wave-vector). 
The linear
response of $M(t)$ to an homogeneous field $H(s)$ applied at time
$s<t$ is given by $\delta \langle M(t)\rangle/\delta
H(s)|_{H=0}=R_{{\bf q}={\bf 0}}(t,s) \equiv \int \D^d x\, R_{\bf x}(t,s)$. 
This quantity, for non-conserved dynamics, is expected to scale as\cite{Cal04b}
\BEQ
R_{{\bf q}={\bf 0}}(t,s) = \widehat{\mathcal A}_R (t-s)^{-(a+1)+d/z}(t/s)^\theta \widehat{\mathcal F}_R(s/t)\;,
\label{eq:scal:Rq}
\EEQ
where the non-universal constant $\widehat{\mathcal A}_R$ is fixed
requiring  $\widehat{\mathcal F}_R(0)=1$ and $\widehat{\mathcal
F}_R(v)$ is a universal scaling function. From Eq.~(\ref{eq:scal:Rx})
one can work out the prediction of LSI for this
susceptibility, finding
\BEQ
\widehat{\mathcal F}^{\rm (LSI)}_R(v)\equiv 1
\label{eq:lsi:q}
\EEQ
and $\widehat{\mathcal A}^{\rm (LSI)}_R = {\mathcal A}_R \int\D^dx\,\Phi({\bf x})$.
In spite of the support to the LSI
prediction Eq.~(\ref{eq:lsi:x}) provided by the result of Monte Carlo
simulations of some critical lattice models, 
recent field-theoretical (FT) computations\cite{Cal02,Cal04b}
suggest the presence of corrections to Eq.~(\ref{eq:lsi:x}).
For the universality class of models in $d$ dimensions with $N$-component
order parameter, short-range interactions with $O(N)$ symmetry, 
and purely dissipative dynamics
(model A of Ref.~\onlinecite{Hoh77}), it
has been predicted
\BEQ
\widehat{\mathcal F}^{\rm (FT)}_R(v) = 1 - \eps^2 \,c_N \Delta f(v) + O(\eps^3)\;,
\label{eq:ft:q}
\EEQ
where $c_N = 3(N+2)/[8(N+8)^2]$, $\eps = 4-d > 0$, 
and $\Delta f(v)\equiv f(0)-f(v) > 0$ ($0\le v\le
1$) is a monotonically increasing and regular 
function of order of unity, whose expression
can be found in Ref.~\onlinecite{Cal02}.
In particular
Eq.~(\ref{eq:ft:q}) with $N=1$ provides, in the dimensional expansion, 
an analytic prediction for the universal scaling function of the
Ising model with Glauber dynamics.
Note that
$\widehat{\mathcal F}^{\rm (FT)}_R = \widehat{\mathcal F}^{\rm
(LSI)}_R + O(\eps^2)$, i.e., the prediction of LSI coincides with the
Gaussian approximation for the scaling function. This observation
accounts for the agreement between LSI and the analytical results for
spherical models with 
$2<d<4$, mentioned above. Indeed, apart from non mean-field
exponents, they typically display Gaussian universal scaling functions.

The discrepancy between $\widehat{\mathcal F}^{\rm (FT)}_R$ and
$\widehat{\mathcal F}^{\rm (LSI)}_R$ casts some doubts on the
effective realization of LSI at the critical point and calls for a
more careful analysis of the results obtained by Monte Carlo simulations,
given that no discrepancy is emerging from the data presented in the past.

\section{The intermediate integrated response}

We consider the susceptibility ($s<t$)
\BEQ
\chi(t,s) = \int_{s/2}^s \!\!\D u \, R_{{\bf q}={\bf 0}}(t,u)
\label{eq:chi}
\EEQ
which is the integrated linear response of the order parameter to a spatially
homogeneous field switched on during the time interval $\left[ s/2, s \right]$.
This integrated response, corresponding to the intermediate protocol proposed
in Ref.~\onlinecite{Hen03}, presents some advantages over the more commonly studied
thermoremanent susceptibility
$\rho_{TRM}(t,s) = \int_0^s \!\!\D u \, R_{{\bf q}={\bf 0}}(t,u)$
and zero-field cooling susceptibility
$\chi_{ZFC}(t,s) = \int_s^t \!\!\D u \, R_{{\bf q}={\bf 0}}(t,u).$
As first noted in Ref.~\onlinecite{Zip00}, $\rho_{TRM}$ is in general hampered by the presence
of a finite-time correction which originates from the response 
to a change in the initial condition. Indeed at the lower integration limit the necessary conditions
for the dynamical scaling regime are not fulfilled and the scaling function (\ref{eq:scal:Rq}) can not be used
in that regime. As discussed in Ref.~\onlinecite{Ple04} a similar
remark applies to $\chi_{ZFC}$ close to the upper integration limit,
which contributes with
a leading term
that is independent of the waiting time $s$, the expected scaling part being only a sub-leading 
correction. These problems with the applicability of the scaling form (\ref{eq:scal:Rq}) are
not encountered for the intermediate integrated response (\ref{eq:chi}).

The expected 
scaling behavior of $\chi(t,s)$ can be worked out from Eq.~(\ref{eq:scal:Rq}):
\BEQ
\chi(t,s) = \widehat{\mathcal A}_\chi t^{-a+d/z} 
\left(\frac{s}{t}\right)^{1-\theta}\widehat {\mathcal F}_\chi(s/t)
\label{eq:skal:chi}
\EEQ
where $\widehat{\mathcal A}_\chi = \widehat{\mathcal A}_R B_\theta$,
\BEQ
\widehat {\mathcal F}_\chi(v) = B_\theta^{-1} \int_{1/2}^1\!\!\D w \,
(1-v w)^{-(a+1)+d/z} w^{-\theta} \widehat {\mathcal F}_R(v w)\;,
\EEQ
and $B_\theta \equiv (1-2^{\theta-1})/(1-\theta)$, so that $\widehat
{\mathcal F}_\chi(0)=1$. 
The prediction of LSI that follows from Eq.~(\ref{eq:lsi:q}) is given
by
\BEQ
\label{eq:lsi:F}
\begin{split}
\widehat {\mathcal F}^{\rm (LSI)}_\chi(v)=& \frac{B_\theta^{-1}}{1-\theta} 
 \big[ \phantom{}_2 F_1(1 - \theta, 1+a-d/z, 2 - \theta; v)\\
&- 2^{\theta-1}\phantom{}_2F_1(1 - \theta, 1+a-d/z, 2 - \theta; v/2)\big]
\end{split}
\EEQ
whereas the FT prediction is obtained from Eq.~(\ref{eq:ft:q}),
%taking into account that $\theta = O(\eps)$, $z = 2 + O(\eps^2)$, and
with $\theta = O(\eps)$, $z = 2 + O(\eps^2)$, and
$\eta = O(\eps^2)$:
\BEQ
\frac{\widehat {\mathcal F}^{\rm (FT)}_\chi(v)}{\widehat {\mathcal
F}^{\rm (LSI)}_\chi(v)}  = 1 - \eps^2 c_N \frac{2}{v}\int_{v/2}^v\!\!\D u\,\Delta f(u) + O(\eps^3)
\label{eq:ft:F}
\EEQ

\begin{figure}[h!]
\centerline{\epsfxsize=2.8in\ \epsfbox{
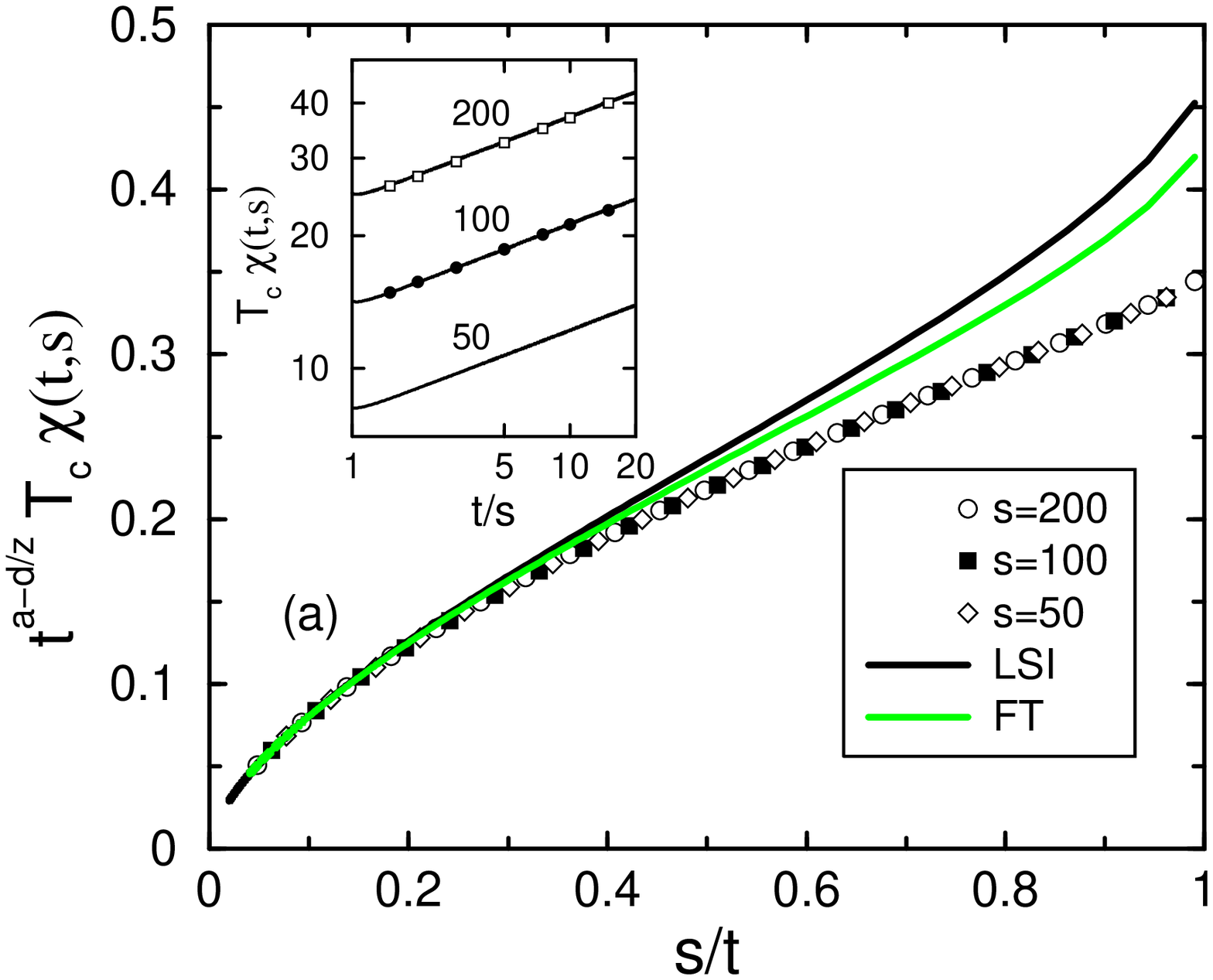}
}
\centerline{\epsfxsize=2.8in\ \epsfbox{
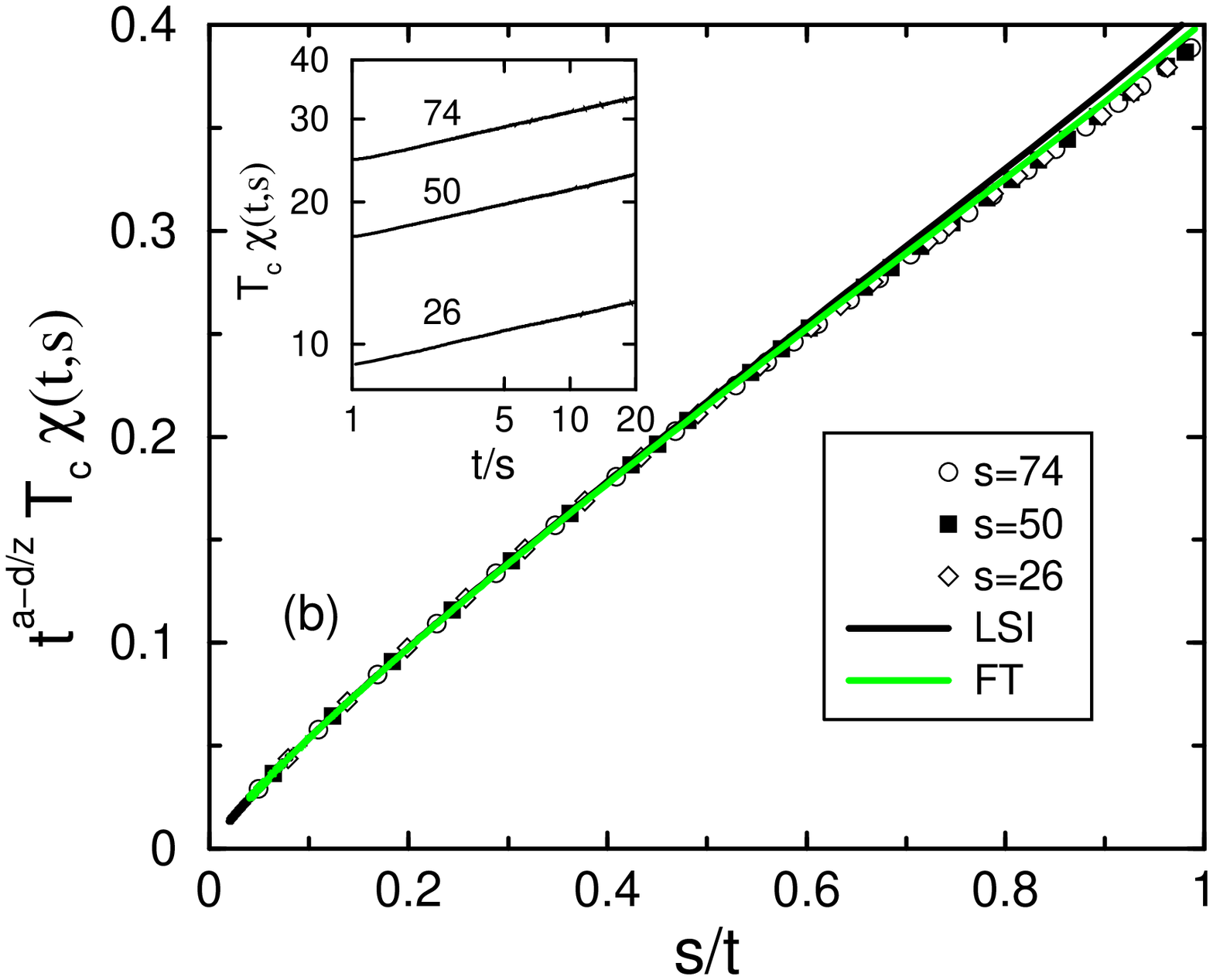}
}
\caption{Rescaled intermediate integrated response vs $s/t$ for 
different waiting times $s$ computed in the critical Ising model
in (a) two and (b) three dimensions. In both cases a perfect data collapse is observed.
The LSI prediction lies systematically above the numerical data
for larger values of $s/t$, whereas the two-loop correction
shifts the theoretical curve closer to the data. Error bars are
smaller than the sizes of the symbols. 
The insets display the power law increase of $\chi$ as a
function of $t/s$ for fixed values of $s$. 
In the inset of (a) we also probe the linear
response regime. 
Solid lines have been obtained with the field strength $H=0.05 J$, whereas
the symbols for $s=200$ and $s=100$ result from simulations with 
$H=0.025 J$ and $H=0.1 J$, respectively. 
The measured susceptibility does not depend
on $H$, %thus 
showing that we are well within the linear 
response regime.
}
\label{fig_1}
\end{figure}

We study the intermediate integrated response (\ref{eq:chi}) by
simulating Ising models on square and on cubic lattices. The systems are thereby prepared in a completely
uncorrelated initial state and then quenched at time $t=0$ to the
critical point. The temporal evolution is realized
using the standard heat-bath algorithm.
At time $t=s/2$ a spatially constant field with strength $H=0.05 J$ ($J$ being the
strength of the nearest neighbor couplings) 
is applied. This external field
is switched off at $t=s$. We checked that for this field strength we are well within the linear
response regime, as shown in the inset of Figure \ref{fig_1}a. The data discussed in the following are free from any finite-size
effects and have been obtained for 
systems consisting of $450\times 450$ spins in two dimensions and
$80\times 80\times 80$ spins in three dimensions. Typically, we averaged over 50000 different
runs with different realizations of the thermal noise.

The temporal evolution of the integrated response 
\BEQ 
\chi(t,s) = \frac{1}{n H}\langle \sum\limits_{i=1}^n S_i(t) \rangle, \quad t> s
\label{eq:chi_num}
\EEQ
is shown in the insets of Figure 1 in two and three dimensions.
Here $S_i(t) \in \{-1,+1\}$ is the value of the Ising spin located at
the lattice site $i$ at time $t$. The sum in (\ref{eq:chi_num}) is over all $n$ sites and
$\langle \ldots \rangle$ indicates the average over the thermal noise.
A similar behavior is observed in two and three dimensions: for a fixed value of the
waiting time $s$ the susceptibility rapidly displays a power-law
increase as a function of $t/s$, with exponent $(d - \lambda)/z$ 
[see Eq.~(\ref{eq:skal:chi})]. Its measured values in two and three
dimensions ($0.186(2)$ and $0.108(2)$, respectively)
are in excellent agreement with
the expected values $0.185$ and $0.108$, 
obtained by using the known values 
of $\lambda$ ($d=2$: $1.60$, $d=3$: $2.78$) and $z$ ($d=2$: $2.17$,
$d=3$: $2.04$).\cite{Hen01,Cal04b}

The scaling behavior (\ref{eq:skal:chi}) is tested in the
main images of Figure 1. Plotting $t^{a-d/z} \, \chi$ as a
function of $s/t$ yields a remarkable data collapse. 
The scaling functions so obtained can be compared with
the available analytical predictions. 
In particular, assuming the well-known values of the critical exponents
appearing in Eq.~(\ref{eq:skal:chi}),\cite{Hen01,Cal04b}
one determines the non-universal constant $\widehat {\mathcal
A}_\chi$ 
from the small $s/t$ behavior of
$t^{a-d/z} \, \chi$.
Then the prediction of LSI and FT are given by $\widehat {\mathcal
A}_\chi (s/t)^{1-\theta}\widehat {\mathcal
F}_\chi^{(LSI),(FT)}$, where $\widehat {\mathcal
F}_\chi^{(LSI)}$ is provided in Eq.~(\ref{eq:lsi:F}) 
with $\theta=0.38$, $a=0.115$ in
$d=2$ and $\theta=0.14$, $a=0.506$ in $d=3$.
The numerical estimate of $\widehat {\mathcal 
F}_\chi^{(FT)}$ is obtained through Eq.~(\ref{eq:ft:F}) with $N=1$,
$\eps=2$ and $1$ in two and three dimensions, respectively,
corresponding to its Pad\'e approximant $[2,0]$, which
does not differ significantly from the approximant $[0,2]$.
The comparison shows that the
LSI curves (black lines) systematically lie above the numerical data
for larger values of $s/t$. Including the two-loop correction
coming from field theory shifts the theoretical curves (grey lines) closer to the data. It has to be stressed
that this is the first time that the existence of corrections to LSI
in critical non-equilibrium systems has been numerically proven.

Finally, we show in Figures 2a and 2b the scaling function  $\widehat
{\mathcal F}_\chi(s/t)$ 
%itself. It is 
obtained from the data of Figure 1 after a multiplication by
$(s/t)^{\theta-1}$ and a proper normalization.
Corrections to the LSI predictions are clearly revealed in this quantity both in two and three
dimensions. Analyzing the scaling function $\widehat {\mathcal F}_\chi(s/t)$ seems therefore to be
the appropriate way for highlighting differences between the theoretical predictions and the
numerical data. In the more common approach where the rescaled susceptibility $s^{a-d/z} \, \chi$
is plotted against $t/s$ corrections are hardly detectable in three dimensions.

\begin{figure}[!h]
\centerline{\epsfxsize=2.8in\ \epsfbox{
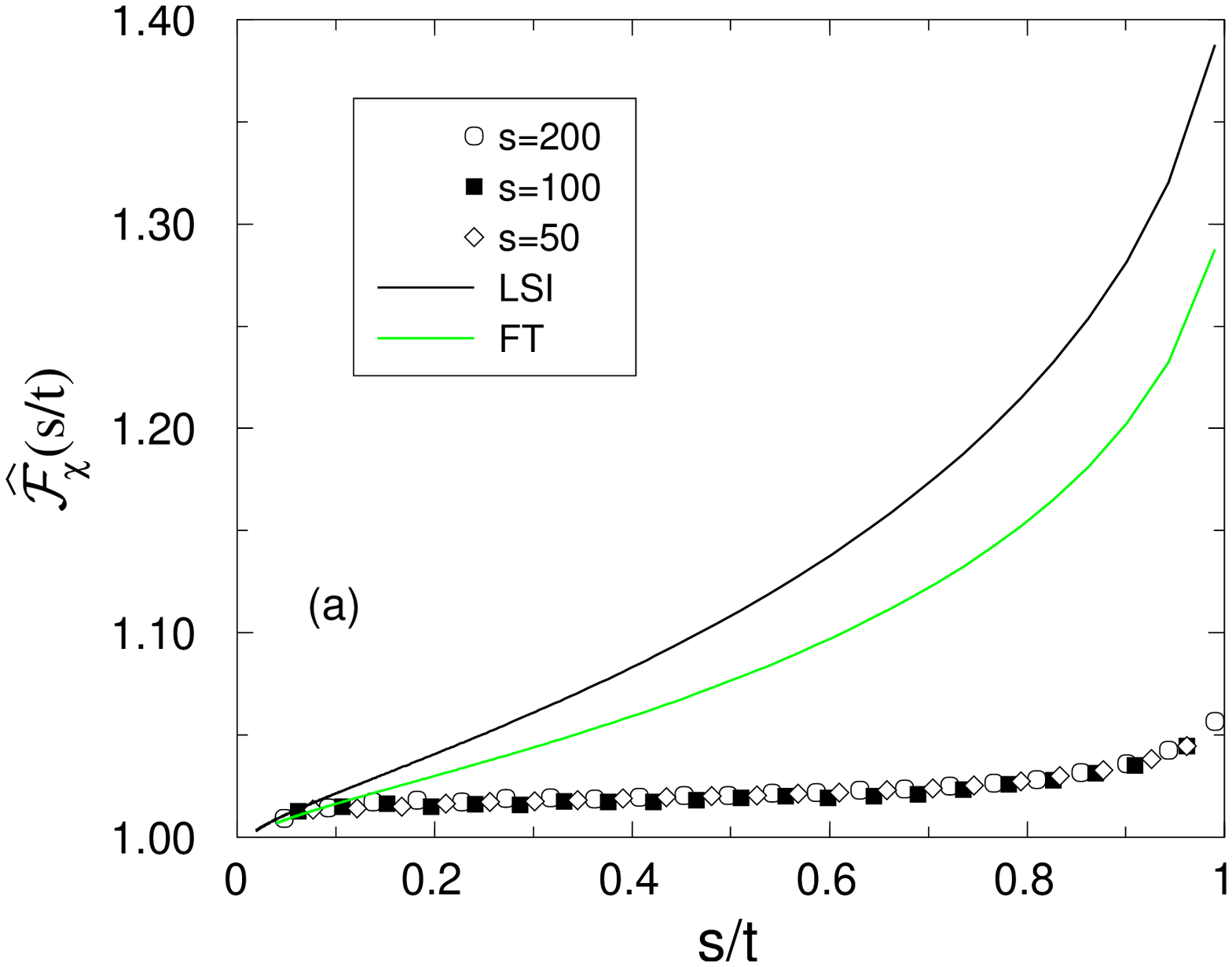}
}
\centerline{\epsfxsize=2.8in\ \epsfbox{
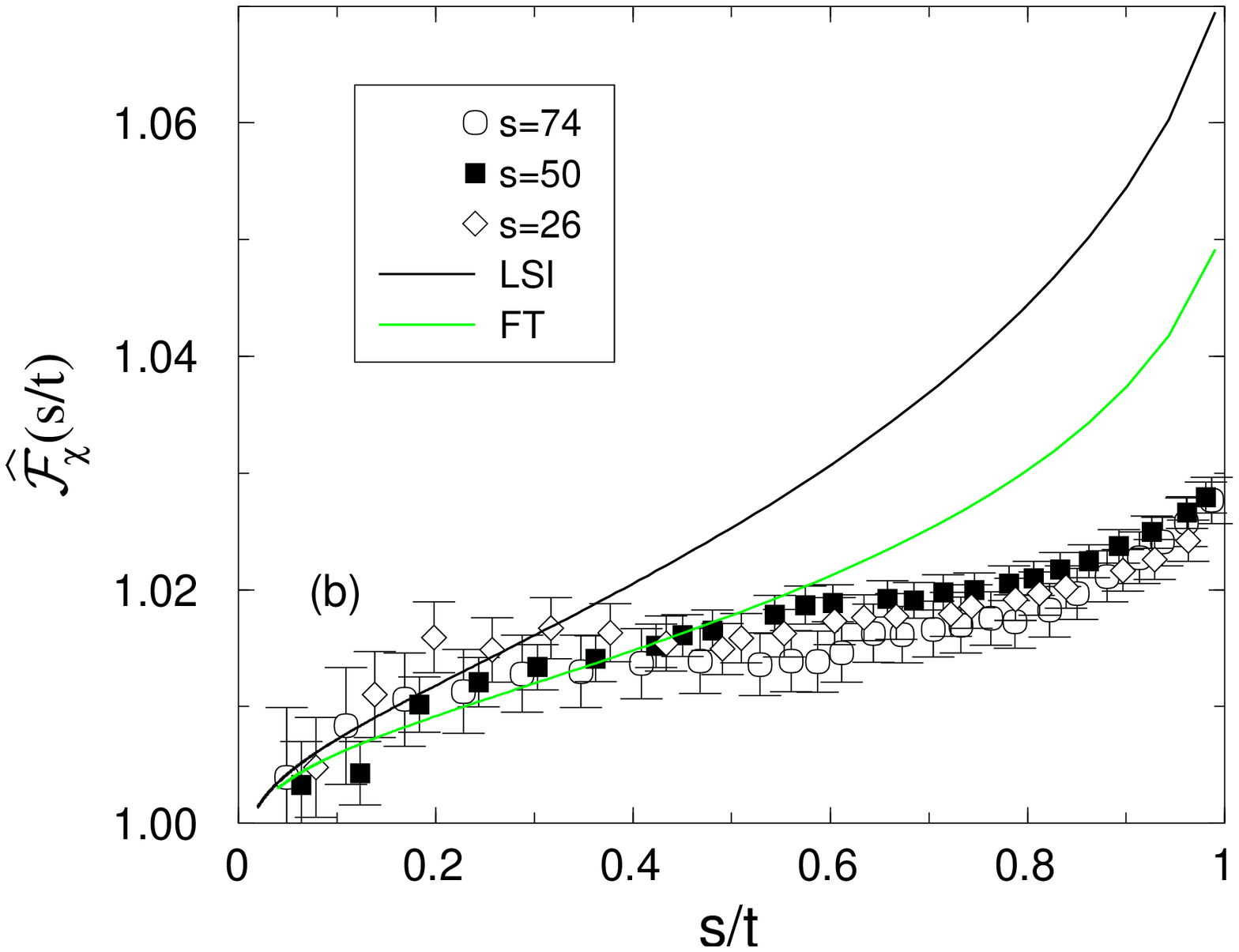}
}
\caption{Comparison of the numerically obtained scaling functions $\widehat {\mathcal F}_\chi$
with the theoretical predictions in (a) two and (b) three dimensions. 
Corrections to LSI are clearly observed for this quantity in both cases. In (a) error bars are 
comparable to the symbol sizes.
}
\label{fig_2}
\end{figure}

\section{Conclusions}
Local Scale Invariance proposes to generalize dynamical scaling to local space- and
time-dependent scale transformations, leading to several
predictions for dynamical correlation and 
response functions. Some of these predictions have been verified in the past through
exact solutions of soluble models and through numerical simulations of non-soluble ones.
In this paper we have focused on the scaling functions of the dynamic response function
in non-equilibrium critical Ising models. 
Previous numerical studies did not reveal 
%in the past 
any
systematic deviations from the LSI prediction when computing local quantities.
Recent field-theoretical calculations, however, yielded a two-loop correction to the LSI
prediction. In the present work we have studied the non-equilibrium response of the total order parameter to
an homogeneous external field, i.e., we have investigated a global quantity.
It has been realized recently that in the usually studied integrated
responses, the thermoremanent susceptibility and the zero-field cooling susceptibility, 
additional terms appear that make the study of the dynamical scaling part (i.e., the aging part)
notoriously difficult. We therefore propose to study the intermediate integrated response
where the field is only switched on during a time interval $[s/2,s]$,
with $0<s<t$, $t$ being the time at which the resulting magnetization
is measured, and $0$ the time at which the quench to the critical
point occurs.
Looking at this quantity we have indeed 
identified corrections to the LSI prediction. 
In addition, we observe that the field-theoretical two-loop correction brings the theoretical
curve closer to the numerical data.
%\vspace{-1cm}
\acknowledgments
We thank Pasquale Calabrese and Malte Henkel for inspiring
discussions. The numerical
work has been done on the IBM supercomputer Jump at the NIC J\"{u}lich
(project Her10). MP acknowledges the support by
the Deutsche Forschungsgemeinschaft through grant no. PL 323/2.

%%%%%%%%%%%%%%%%%%%%%%%%%%%%%%%%%%%%%%%%%%%%%%%%%%%%%%%%%%%%%%%%%%%%%%%%%%%%%%%%


\begin{references}
\bibitem{footnote1} Note that
LSI does not fix the
values of the critical exponents appearing in Eq.~(\ref{eq:scal:autoR}).
\end{references}
\end{document}